\documentstyle[art8,aaspp4,flushrt,tighten,psfig]{article}

\slugcomment{Astrophysical Journal Letters, in press}

\lefthead{Feldmeier \& Shlosman}
\righthead{Wind evolution towards critical and overloaded solutions}

\newcommand{\pa}{\partial}
\newcommand{\cri}{_{\rm c}}
\newcommand{\ter}{_{\rm t}}
\newcommand{\Abb}{_{\rm A}}
\newcommand{\Abbcri}{_{\rm Ac}}
\newcommand{\Abbter}{_{\rm At}}

\begin{document}

\title{RUNAWAY OF LINE-DRIVEN WINDS TOWARDS CRITICAL AND OVERLOADED
SOLUTIONS}

\author{Achim Feldmeier}

\affil{Imperial College, Prince Consort Road, London SW7 2BZ, England
\\ email: {\tt a.feldmeier@ic.ac.uk}}

\and

\author{Isaac Shlosman}

\affil{University of Kentucky, Lexington, KY 40506, U.S.A. \\ email:
{\tt shlosman@pa.uky.edu}}

\begin{abstract}

Line-driven winds from hot stars and accretion disks are thought to
adopt a unique, critical solution which corresponds to maximum mass
loss rate and a particular velocity law. We show that in the
presence of negative velocity gradients, radiative-acoustic (Abbott)
waves can drive shallow wind solutions towards larger velocities and
mass loss rates. Perturbations introduced downstream from the wind
critical point lead to convergence towards the critical solution. By
contrast, low-lying perturbations cause evolution towards a
mass-overloaded solution, developing a broad deceleration region in
the wind. Such a wind differs fundamentally from the critical
solution. For sufficiently deep-seated perturbations, overloaded
solutions become time-dependent and develop shocks and shells.

\end{abstract}

\keywords{accretion disks --- galaxies: active --- cataclysmic
variables --- hydrodynamics --- stars: mass loss --- stars: winds}
\twocolumn
 
\section{Introduction}

Atmospheres of hot luminous stars and accretion disks in active
galactic nuclei and cataclysmic variables form extensive outflows due
to super-Eddington radiation fluxes in UV resonance and subordinate
lines. An understanding of these winds is hampered by the pathological
dependence of the driving force on the flow velocity gradient. Castor,
Abbott, \& Klein (1975; CAK hereafter) found that line-driven winds
(hereafter LDWs) from O~stars should adopt a unique, critical state
which corresponds to maximum mass loss rate. The equation of motion
for a 1-D, spherically symmetric, polytropic outflow subject to a
Sobolev line force allows for two infinite families of so-called
shallow and steep solutions. However, none of these families can
provide for a {\it global} solution alone. Shallow solutions do not
reach infinity, while steep solutions do not extend into the subsonic
regime including the photosphere. The critical wind starts then as the
fastest shallow solution and switches at the critical point in a
continuous and differentiable manner to the slowest steep
solution. Hence the critical point and not the sonic point determines
the bottleneck in the wind. This description in principle applies
equally to winds from stars and accretion disks.

A physical interpretation of the CAK critical point was given by
Abbott (1980), who derived a new type of radiative-acoustic waves
(hereafter Abbott waves). These waves can propagate inward, in the
stellar rest frame, only from below the CAK critical point. Above the
critical point, they are advected outwards. Hence, the CAK critical
point serves as an information barrier, much as the sonic or Alfv\'en
points in thermal and hydromagnetic winds. Abbott's analysis was
challenged by Owocki \& Rybicki (1986) who found for a pure absorption
LDW the signal speed to be the sound speed and not the much faster
Abbott speed. As noted already by these authors, this should be a
consequence of assuming pure line absorption, which does not allow for
any radiatively modified, inward wave mode. Meanwhile there is ample
evidence for Abbott waves in time-dependent wind simulations (Owocki
\& Puls 1999).

Shallow solutions fail to reach infinity because they cannot perform
the required spherical expansion work, implying that the flow starts
to decelerate. Since this usually occurs very far out in the wind, the
local wind speed is much larger than the local escape speed, and the
wind escapes to infinity. Thus, a simple generalization of the CAK
model allowing for flow deceleration renders shallow solutions
globally admissible. This raises a fundamental question of why the
wind would adopt the critical solution at all, and attain the critical
mass loss rate and velocity law, as proposed by CAK.

In this Letter we analyze a physical mechanism which drives shallow
solutions towards the critical one, and discuss under what conditions
this evolutions does not terminate at the CAK solution, but continues
into the realm of overloaded solutions. We find that simulations so
far were affected by numerical runaway towards the critical solution,
by not accounting for Abbott waves in the Courant time step.

\section{Abbott waves}

Abbott waves are readily derived by bringing the wind equations into
characteristic form. We consider a 1-D planar wind of velocity
$v(z,t)$ and density $\rho(z,t)$, assuming zero sound speed. The
continuity and Euler equation are,
\begin{equation}
 \label{continuity}
 {\pa \rho \over \pa t} + v {\pa \rho \over \pa z} + \rho {\pa v
 \over \pa z} = 0,\\
\end{equation}
\begin{equation}
\label{euler} 
E \equiv {\pa v \over \pa t} + v\, {\pa v \over \pa z} + g(z) -
 C \, F(z) \left({\pa v / \pa z \over \rho}\right)^\alpha = 0.
  \end{equation}
 Here, $g(z)$ and $F(z)$ are gravity and radiative flux,
respectively. The CAK line force is given by $g_{\rm l} \equiv C \,
F(z) \, (v' / \rho)^\alpha$ (with $v' \equiv \partial v/\partial z$),
with constant $C$ and exponent $0<\alpha <1$. The unique, stationary
CAK wind, $v\cri(z), \rho\cri(z)$, is found by requiring a critical
point at some $z\cri$. The number of solutions for $vv'(z)$ changes
from 2 to 1 at $z\cri$ (which is a saddle point), hence $\partial
E/\partial (vv')\cri=0$ holds. Writing $C$ in terms of critical point
quantities, the Euler equation becomes,
 
 $${\pa v \over \pa t} + v\, {\pa v \over \pa z} + g(z) - $$
\begin{equation}
\alpha^{-\alpha} (1-\alpha)^{-(1-\alpha)} \, {F(z) \over F(z\cri)} \;
g(z\cri)^{1-\alpha} (\rho\cri v\cri)^\alpha \left({\pa v / \pa z \over
\rho}\right)^\alpha =0.
 \end{equation}
 \noindent Note that for stationary planar winds, $\rho v$ is
constant. If, in addition, $g$ and $F$ are taken constant with height,
and $\rho\cri v\cri \, v'/\rho$ is replaced by $vv'/\dot m$, with
normalized mass loss rate $\dot m \equiv \rho v/\rho\cri v\cri$, one
finds that $E$ does no longer depend explicitly on $z$ for stationary
solutions. Hence, $vv'$ is independent of $z$, too. This implies that
$z\cri$ is ill-defined, and every point of the CAK solution is a
critical point. CAK removed this degeneracy by introducing gas
pressure terms. Here we take a different approach and assume
$g=z/(1+z^2)$. A situation with roughly constant radiative flux and
gravity showing a maximum at finite height could be encountered above
isothermal disks around compact objects (cf. Feldmeier \& Shlosman 1999). The
critical point is determined by the regularity condition, $dE/dz\cri=0$, hence
$z\cri=1$ and the critical point coincides with the gravity maximum. For
simplicity also we chose $\alpha=1/2$ from now on, which is reasonably
close to realistic values $\alpha \le 2/3$ (Puls et al.~1999). None of
our results should depend qualitatively on the assumptions made so
far. The Euler equation is
 \begin{equation}
 \label{euler2} 
 {\pa v \over \pa t} + v\, {\pa v \over \pa z} + g(z) -
 2\sqrt{g\cri \rho\cri v\cri} \; \sqrt{\pa v / \pa z \over \rho} =0,
 \end{equation}
 \noindent where $g\cri \equiv g(z\cri)$. The stationary solutions for
wind acceleration are given by
 \begin{equation}
 \label{windsol}
 vv'(z) = {g\cri \over \dot m} \, \left( 1 \pm \sqrt{1- {\dot mg(z)
\over g\cri}}\right)^2,
 \end{equation}
 \noindent where plus and minus signs refer to steep and shallow
solutions, respectively. For $\dot m \le 1$, shallow and steep
solutions are globally, i.e., everywhere, defined. For $\dot m>1$,
solutions are called {\it overloaded}, and become imaginary in a
neighborhood of the gravity maximum. These winds carry too large mass
loss rates and eventually stagnate.

Next we put the Euler equation into quasi-linear form, which does {\it
not} mean to linearize it. Differentiating $E$ with respect to $z$
(Courant \& Hilbert 1962; Abbott 1980) and introducing $f\equiv \pa
v/\pa z$, eqs.~(\ref{continuity}, \ref{euler2}) become,
 \begin{eqnarray}
 \label{characon}
 &&\left[{\pa \over \pa t} + v \, {\pa \over \pa z}\right] \rho \;+\;
 \rho f = 0,\\
 \label{charaeul}
 &&\left[{\pa \over \pa t} + (v+v\Abb) \, {\pa \over \pa
 z} \right] {f \over \rho} \;+\; {1\over \rho}\, {\pa g \over \pa z} =
 0,
 \end{eqnarray}
 \noindent with inward Abbott speed in the rest frame, $v\Abb \equiv
-\sqrt{g\cri v/\dot m v'}$. In the WKB approximation, individual
spatial and temporal variations are much larger than the inhomogeneous
term $g'/\rho$ in eq.~(\ref{charaeul}), and the latter can be
neglected. Consequently, $v'/\rho$ is a Riemann invariant propagating
at characteristic speed $v+v\Abb$.  Perturbations of $v'/\rho$
correspond to the amplitude of a wave propagating at phase speed
$v+v\Abb$. Note that $v'/\rho$ is proportional to the Sobolev line
optical depth, indicating that this wave is a true radiative mode.

The second characteristic is determined by the continuity equation
(\ref{characon}). In the advection operator in square brackets, $v$
has to be read as $v+0$ in the zero-sound speed limit. This outwards
propagating invariant corresponds to a sound wave, with amplitude
$\rho$ scaling with gas pressure.

At the critical point, $\dot m=1$ and $vv'(z\cri)=g(z\cri)$ after
eq.~(\ref{windsol}), hence $v\Abbcri=-v\cri$ (where we introduced
$v\Abbcri \equiv v\Abb(z\cri)$). Abbott waves stagnate at the critical
point, in analogy with sound waves at the sonic point. For shallow
solutions, $\dot m<1$ and $vv'<v\cri v\cri'$ from eq.~(\ref{windsol}),
hence $v+v\Abb <0$. Shallow LDW solutions are therefore the
subcritical analog to solar wind breezes.

Because in the rest frame, the inward Abbott mode can propagate at
larger absolute speeds than the outward sound mode, Abbott waves can
determine the Courant time step in time-explicit hydrodynamic
simulations. Violating the Courant step results in numerical
instability. Despite this fact, Abbott waves along shallow solutions
were never considered in the literature.

\section{Wind convergence towards the critical solution}
\label{acceleration}

We turn our attention to a physical mechanism which can drive LDWs
away from shallow solutions, and towards the critical one. Starting
from an arbitrary shallow solution as initial condition, we explicitly
introduce perturbations at some fixed location in the wind and study
their evolution. In order to keep {\it unperturbed} shallow solutions
stable in numerical simulations, we fix one outer boundary condition,
according to inward propagating Abbott waves. Either a constant mass
loss rate at the outer boundary or non-reflecting boundary conditions
(Hedstrom 1977) serve this aim. At the inner, subcritical boundary, we
also fix one boundary condition, according to incoming sound waves.
Non-reflecting boundary conditions and $\rho=const$ give similar
results.

Wind convergence towards the critical solution is then triggered by
negative flow velocity gradients. Allowing for $v'<0$ turns the inward
Abbott mode of phase speed $v+v\Abb <0$ in the rest frame into an
outwards propagating mode. This is readily seen for a line force which
is zero for negative $v'$, i.e., when all photons are absorbed at a
resonance location between the photosphere and the wind point. The
Euler equation simplifies to that for an ordinary gas, with
characteristic speed $v-0 >0$ in the zero sound speed limit. At the
other extreme, for a purely local line force where the unattenuated
stellar or disk radiation field reaches the wind point, $g_{\rm l}
\propto \sqrt{|v'|}$. Here the Abbott phase speed is found to be
$v+v\Abb$, with $v\Abb = +\sqrt{ -g\cri v/\dot m v'}$ for $v'<0$.

Consider then a sawtooth-like velocity perturbation (sinusoidal
perturbation lead to similar results). Slopes $v'>0$ propagate
inwards, slopes $v'<0$ propagate outwards. Hence, as a kinematical
consequence, a sawtooth which is initially symmetric with respect to
the underlying stationary velocity law evolves towards larger
velocities. This is demonstrated in Figure~1 where, in course of time,
a periodic sawtooth perturbation is introduced at $z=2$. The line
force is assumed to be $\propto \sqrt{|v'|}$, and the initial shallow
solution has $\dot m=0.8$. The figure shows 2${1 \over 2}$
perturbation cycles. For upward pointing kinks the slopes propagate
apart and a flat velocity law develops between them.  At each time
step $dt$, a new increment $dv = 4dt\, \delta v/T$ ($\delta v$ and $T$
being the amplitude and period of the sawtooth) is added at $z=2$,
hence the flattening velocity law does not show up in region A of
Figure~1. Overall, the wind speed at the perturbation site evolves
towards larger values during these phases. On the other hand, for
downward pointing kinks of the sawtooth, $-\delta v$, the two
approaching slopes merge, and the wind speed evolves back towards its
unperturbed value after each decrement $-dv = -4dt\, \delta v/T$. The
wind velocity hardly evolves during these phases, cf.~region B of
Figure~1. Over a full perturbation cycle, the wind speed clearly
increases.  

\begin{figure}[ht]
\vbox to3.6in{\rule{0pt}{3.6in}} 
\includegraphics{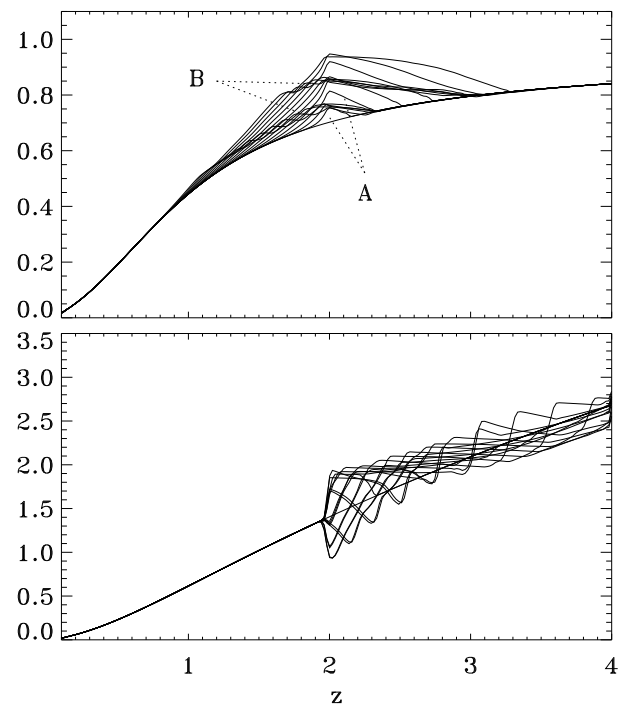}    
\caption{{\it Upper panel\/}: Evolution of a shallow wind
during 2$1 \over 2$ periods of a sawtooth perturbation of amplitude
10\% being introduced at $z=2$. Regions `A' and `B' correspond to
phases where upward and downward pointing kinks, respectively, are
introduced into the flow. {\it Lower panel\/}: stable Abbott wave
excitation in the critical wind, at 35\% perturbation amplitude.}
\end{figure}     

Essentially any perturbation which introduces negative $v'$ will
accelerate the wind. The amplitude of the perturbation is rather
irrelevant since, with decreasing perturbation wavelength, negative
$v'$ occur at ever smaller amplitudes. However, in more realistic
winds, dissipative effects may smear out short-scale perturbations 
before they can grow. Details of the physical mechanism will be
discussed elsewhere.

{\it If} the perturbation lies downstream from the critical point, the
wind converges to the critical solution. Namely, as soon as the
perturbation site comes to lie on the supercritical part of the CAK
solution during its evolution, positive velocity slopes propagate {\it
outwards}, and combine with negative slopes to a full wave train. No
information is propagated upstream. This unconditional stability of
the outer CAK solution is shown in the lower panel of Fig.~1.

\section{Wind convergence towards overloaded solutions}
\label{overloaded}

Wind runaway towards larger speeds as caused by perturbations
introduced {\it upstream} from the critical point does not terminate
at the critical CAK solution. For low-lying perturbations,
communication with the wind base is still possible once the
subcritical branch of the CAK solution is reached.  The wind gets
further accelerated into the domain of mass-overloaded solutions
(where $vv'> v\cri v'\cri$ and hence $v> v\cri$ for $z<z\cri$
according to eq.~\ref{windsol}), until a generalized critical point
develops, which prevents inward propagation of Abbott waves and
adjustment of the mass loss rate. Such generalized critical points are
given by `termination' points, $z\ter$, of overloaded solutions, where
the velocity becomes imaginary. At $z\ter$, the number of real
solutions $vv'(r)$ changes from 2 (shallow and steep) to 0. Hence,
termination points are defined by the same condition as the CAK
critical point (at which the number of solutions changes from 2 via 1
to 2), $\pa E/\pa (vv')\ter=0$. From the stationary version of
eq.~(\ref{euler2}), $v\Abbter =-v\ter$, hence Abbott waves stagnate at
termination points, and the latter are generalized critical points.

The fact that perturbations with negative $v'$ accelerate the wind
either to the critical or an overloaded state can be casted into
black-hole conjecture (Penrose 1965): a LDW avoids a `naked' base, and
encloses it with a critical surface.

Since to each $z\ter$ there corresponds a unique, supercritical mass
loss rate, the latter is determined by the perturbation {\it location}
alone. Using $v\Abbter=-v\ter$, one finds $\dot m\ter= g\cri/g\ter >1$
for a planar wind with constant radiative flux.

At a termination point, $vv'$ jumps to the decelerating branch, $vv'
<0$. Beyond a well-defined location above the gravity maximum, the
super-CAK mass loss rate can again be lifted by the line force, and
$vv'$ jumps back to the accelerating branch. Hence, two stationary
kinks occur in the velocity law. Figure~2 shows a hydrodynamic
simulation of the evolution towards an overloaded solution.
Sawtooth-type velocity perturbations were introduced at $z=0.8$.
Correspondingly, $\dot m=1.025$ for the overloaded solution, using
$g=z/(1+z^2)$.

\begin{figure}[ht]
\plotone{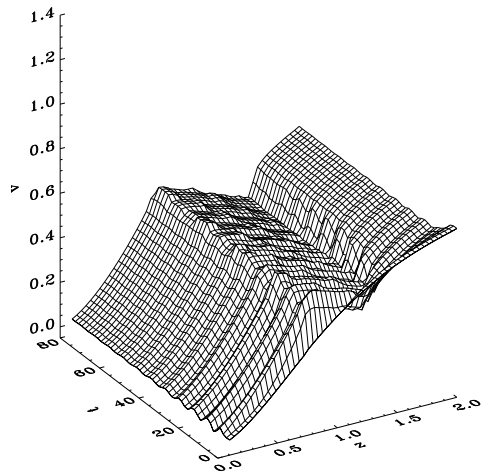}  
\caption{Wind evolution towards a stationary, overloaded
solution showing an extended decelerating region. A periodic sawtooth
perturbation is introduced into a shallow solution at $z=0.8$,
upstream from the critical point at $z\cri=1$.  }
\end{figure}

Future work has to clarify whether LDWs show deep-seated
perturbations. It seems however unlikely that they would occur at a
unique location. Hence, overloaded winds should be non-stationary and
show a {\it range} of supercritical mass loss rates.

More fundamentally, time-dependent overloaded solutions occur already
for single, unique perturbation sites, once the latter lie below a
certain height. For the present wind model, this is at $z \approx
0.66$. The overloading is then so severe and the decelerating region
so broad that negative wind speeds result (cf.~Poe, Owocki, \& Castor
1990). The corresponding mass loss rates are still only a few percent
larger than the CAK value. The gas which falls back towards the
photosphere collides with outflowing gas, and a time-dependent
situation develops. Within each perturbation period, a shock forms in
the velocity law, supplemented by a dense shell. These shocks and
shells propagate outwards (Feldmeier \& Shlosman 2000).

Although strong perturbations introducing negative velocity gradients
can appear already in O~star winds, accretion disk winds are the prime
suspects. The reason for this is that accretion processes and their radiation
fields in cataclysmic variables and galactic nuclei are intrinsically
variable on a range of timescales (Frank, King, \& Raine 1992), and
that disk LDWs are driven by a combination of uncorrelated, local and
central radiation fluxes.
 
\section{Summary}

We find that shallow solutions to line-driven winds are subcritical
with respect to Abbott waves (sub-abbottic). These waves cause
shallow solutions to evolve towards larger speeds and mass loss rates
because of the asymmetry of the line force with regard to positive and
negative velocity gradients and because perturbations with opposite signs of
$dv/dz$ propagate in opposite directions. Steep velocity slopes
propagate towards the wind base, steepen the inner wind and lift
it to higher mass loss rates.   In the presence of enduring wind
perturbations,  this proceeds until a critical point forms and
Abbott waves can no longer penetrate inwards.

The resulting solution does not necessarily correspond to the CAK
wind. For perturbations which originate below the critical point, the
developing Abbott wave barrier is found to be the termination point of
a mass-overloaded solution. The velocity law acquires a kink at the
termination point, where the wind starts to decelerate. Whether the
wind converges to a critical or overloaded solution depends entirely
on the {\it location} of perturbations, and not, e.g., on boundary
conditions at the wind base.

If Abbott waves are not accounted for in the Courant time step of
hydrodynamic simulations, we find that {\it numerical} runaway can
drive the solution towards the critical CAK wind. A detailed
discussion of this will be given elsewhere.

Future work has to clarify whether and where perturbations causing
local flow deceleration, $dv/dz<0$, can occur in LDWs. Overloaded
winds may be detected observationally. While their mass loss rates
should still be close to CAK values, broad regions of decelerating
flow could be identified in P~Cygni line profiles. Furthermore, shocks
occurring in overloaded solutions with infalling gas may contribute to
the X-ray emission from LDWs, besides shocks from the line-driven
instability (Lucy 1982; Owocki et al.~1988). Note that the present wind
runaway occurs already in the lowest order Sobolev approximation,
and is therefore, unrelated to the line-driven instability which
depends on velocity curvature terms (Feldmeier 1998).

\acknowledgments

We thank R.~Buchler, J.~Drew, R.~Kudritzki, C.~Norman, S.~Owocki, and
J.~Puls for intense blackboard discussions, and the referee, Stan
Owocki, for suggestions improving the manuscript. This work was
supported in part by PPA/G/S/1997/00285, NAG 5-3841, WKU-522762-98-6
and HST~GO-08123.01-97A.

\end{document}